\documentclass[twocolumn,aps,pra,showpacs,amsmath,superscriptaddress]{revtex4}
\usepackage{amssymb}
\usepackage{mathrsfs}
\usepackage{graphicx}
\usepackage{float}

\begin{document}

\title{Ground-state and dynamical properties of hard-core bosons on
one-dimensional incommensurate optical lattices with harmonic trap}

\author{Xiaoming Cai}
\affiliation{Beijing National
Laboratory for Condensed Matter Physics, Institute of Physics,
Chinese Academy of Sciences, Beijing 100190, China}
\author{Shu Chen}
\affiliation{Beijing National Laboratory for Condensed Matter
Physics, Institute of Physics, Chinese Academy of Sciences, Beijing
100190, China}
\author{Yupeng Wang}
\affiliation{Beijing National Laboratory for Condensed Matter
Physics, Institute of Physics, Chinese Academy of Sciences, Beijing
100190, China}
\date{ \today}

\begin{abstract}
We study properties of the strongly repulsive Bose gas on
one-dimensional incommensurate optical lattices with a harmonic
trap, which can be deal with by using the exact numerical method
through the Bose-Fermi mapping. We first exploit the phase
transition of the hard-core bosons in the optical lattices from
superfluid-to-Bose-glass phase as the strength of the incommensurate
potential increases. Then we study the dynamical properties of the
system after suddenly switching off the harmonic trap. We calculate
the one-particle density matrices, momentum distributions, the
natural orbitals and their occupations for both the static and
dynamic systems. Our results indicate that the Bose-glass phase and
superfluid phase display quite different properties and expansion
dynamics.
\end{abstract}

\pacs{
 05.30.Rt, 
 05.30.Jp, 
 72.15.Rn  
 }

\maketitle

\section{Introduction}

Recently, the technical advances in cold atom trapping allowed the
experimental realization of the Anderson localization
\cite{Anderson} in quantum matter waves \cite{Billy,Roati}. The good
tunability and controllability of optical lattices offer myriad
opportunities for studying the disorder effects in ultracold atom
system \cite{Lewenstein}. So far, different techniques have been
devised for the introduction of disorder in the ultracold atom
system, such as speckle filed patterns in the optical lattice
\cite{Billy,Lye}, random localized impurities by loading a mixture
of two kinds of atoms with heavy and light masses \cite{Gavish}, and
incommensurate bichromatic optical lattices by superimposing two
one-dimensional (1D) optical lattices with incommensurate frequency
\cite{Roati,Fallani}. In particular, the experiment \cite{Fallani}
has provided evidences of the existence of a Bose-glass (BG) phase
\cite{Fisher89}.

Interesting phenomenons are expected to appear in disordered systems
when the interplay of disorder and interactions is taken into
account. Interactions between atoms can be controllably tuned by
Feshbach resonances in ultra-cold atom systems. While disorder can
lead to localization of the wave function of a particle,
delocalization can arise as the consequence of interactions in some
many-body systems. Theoretically, for a repulsive Bose gas it has
been predicted that there is a quantum phase transition from a
superfluid phase to an insulating BG phase with localized
single-particle states as disorder is increased
\cite{Giamarchi,Fisher,Delande,Fontanesi,Egger,Egger2,Cai}, however
unambiguous observation of the superfluid-Bose-glass transition is
still under debate \cite{Damski,Delande}. A lot of attention
\cite{Gurarie,Roux,Deng,Roscilde,Orso} has been payed to
investigating the combined role of disorder and interactions in the
strongly interacting ultra-cold atomic system. Except numerical or
approximate approaches, the exact solution for the many-body systems
with the interplay of disorder and interaction are rarely known. In
this paper, we study the interacting bosons on the incommensurate
optical lattices with harmonic trap in the limit case with
infinitely repulsive interaction which can be solved exactly. The 1D
Bose gas with infinitely repulsive interaction is known as the
hard-core boson (HCB) or Tonks-Girardeau (TG) gas \cite{Girardeau},
which can be exactly solved via the Bose-Fermi mapping
\cite{Girardeau} and has attracted intensive theoretical attention
\cite{Girardeau1,Minguzzi,Gangardt}. Experimental access to the
required parameter regime has made the TG gas a physical reality
\cite{Paredes,Kinoshita}.  For the HCB in 1D optical lattices, it is
convenient to use the exact numerical approach proposed by Rigol and
Muramatsu \cite{Rigol}. Following the exact numerical approach, we
calculate the static properties of the hard-core bosons, such as
one-particle density matrices, density profiles, momentum
distributions, natural orbitals and their occupations, to exploit
the superfluid-to-BG phase transition for the systems in
incommensurate optical lattices with harmonic confining trap.
Furthermore, we study the nonequilibrium dynamical properties of
expanding clouds of hard-core bosons on the 1D incommensurate
lattices after turning off the harmonic trap suddenly. We find that
the expansion dynamics for the superfluid phase and BG phase exhibit
quite different behaviors, which may serve as a signature for
experimentally detecting the transition from superfluid-to-BG phase.

The paper is organized as follows. In Section II, we present the
model and the exact approach used in this paper. In Section III, we
show properties of the ground-state for hard-core bosons on the
incommensurate optical lattice with a harmonic trapping potential.
Section IV is devoted to studying the nonequilibrium dynamics of the
system after the harmonic trap is suddenly switched off. Finally, a
summary is presented in Section V.

\section{Model And Method}

In the present section we describe the exact approach which we used
to study 1D hard-core bosons on the incommensurate lattice with an
additional harmonic trap. Under the single-band tight-binding
approximation, the system of $N$ hard-core bosons in the 1D optical
lattice can be described by the following Hamiltonian:
\begin{equation}
\label{eqn2}
H=-t\sum_i(b^\dagger_ib_{i+1}+\mathrm{H.c.})+\sum_iV_in^b_i,
\end{equation}
where $b^\dagger_i(b_i)$ is the creation (annihilation) operator of
the boson which fulfills the hard-core constraints \cite{Rigol},
{\it i.e.,}  the on-site anticommutation $(\{b_i,b^\dagger_i\}=1)$
and $[b_i,b^\dagger_j]=0$ for $i\neq j$; $n^b_i$ is the bosonic
particle number operator; $t$ is the hopping amplitude set to be
the unit of the energy $(t=1)$; $V_i$ is given by
\begin{equation}
V_i=V_I\mathrm{cos}(\alpha2\pi i+\delta)+V_H(i-i_0)^2.
\end{equation}
Here $V_I$ is the strength of incommensurate potential with $\alpha$
as an irrational number characterizing the degree of the
incommensurability and $\delta$ an arbitrary phase (in our
calculation it is chosen to be zero for convenience, without loss of
generality), $V_H$ is the strength of harmonic trap and $i_0$ is the
position of the vale of the harmonic trap.

In order to calculate the properties of hard-core bosons, it is
convenient to use the Jordan-Wigner transformation \cite{Jordan}
(JWT) or Bose-Fermi mapping for the lattice model
\begin{equation}
b^\dagger_j=f^\dagger_j\prod^{j-1}_{\beta=1}e^{-i\pi f^\dagger_\beta
f_\beta},b_j=\prod^{j-1}_{\beta=1}e^{+i\pi f^\dagger_\beta
f_\beta}f_j,
\end{equation}
which maps the Hamiltonian of hard-core bosons into the Hamiltonian
of noninteracting spinless fermions
\begin{equation}
\label{eqn1}
H_F=-\sum_i(f^\dagger_if_{i+1}+\mathrm{H.c.})+\sum_iV_in^f_i ,
\end{equation}
where $f^\dagger_i(f_i)$ is the creation (annihilation) operator of
the spinless fermion and $n^f_i$ is the particle number operator.
The ground-state wave function of the system with $N$ spinless free
fermions can be obtained by diagonalizing Eq.(\ref{eqn1}) and can be
represented as
\begin{equation}
\label{eqn3}
|\Psi^G_F\rangle=\prod^N_{n=1}\sum^L_{i=1}P_{in}f^\dagger_i|0\rangle
,
\end{equation}
where $L$ is the number of the lattice sites, $N$ is the number of
fermions (same as bosons), and the coefficients $P_{in}$ are the
amplitude of the $n$-th single-particle eigenfunction at the $i$-th
site which can form an $L \times N$ matrix $P$ \cite{Rigol}.

In order to get the static properties of the ground-state, we
calculate the one-particle Green function for the hard-core bosons
defined by
\begin{equation}
G_{ij}=\langle\Psi^G_{HCB}|b_ib^\dagger_j|\Psi^G_{HCB}\rangle\\
= \langle\Psi^A|\Psi^B\rangle ,
\end{equation}
where $|\Psi^G_{HCB}\rangle$ is the ground-state of hard-core
bosons, and $\langle\Psi^A| =
\left(f^\dagger_i\prod_{\beta=1}^{i-1}e^{-i\pi f^\dagger_\beta
f_\beta}|\Psi^G_F\rangle\right)^\dagger$, $ |\Psi^B\rangle
=f^\dagger_j\prod_{\gamma=1}^{j-1}e^{-i\pi f^\dagger_\gamma
f_\gamma}|\Psi^G_F\rangle $. Explicitly the state $\left| \Psi
^A\right\rangle $ can be represented as $ \left| \Psi^A\right\rangle
=\prod_{n=1}^{N+1}\sum_{l=1}^LP_{ln}^{A}f_l^{\dagger }\left|
0\right\rangle $ with $ P_{ln}^{A} = - P_{ln} $ for $l\leq i-1$,
$P_{ln}^{A} =P_{ln}$ for $l\geq i$ with $n\leq N$, $P_{iN+1}^{A}=1$
and $P_{lN+1}^{A}=0$ $\left( l\neq i\right) $. Similarly we can get
$P^{B}$ for the state $|\Psi^B\rangle$ with the replacement of $i$
by $j$. The Green function is a determinant dependent on the
$L\times \left( N+1\right) $ matrices $P^{A}$ and $P^{B}$
\cite{Rigol}
\begin{eqnarray}
G_{ij}=\left\langle \Psi^A|\Psi^B\right\rangle =\det \left[ \left(
P^{A}\right)^\dagger P^{B}\right] .
\end{eqnarray}
The one-particle density matrix can be evaluated from the relation
\begin{equation}
\rho_{ij}=\langle
b^\dagger_ib_j\rangle=G_{ij}+\delta_{ij}(1-2G_{ii}).
\end{equation}
Alternatively one can use the method proposed by Pareder {\it et
al.} \cite{Paredes} to calculate the one-particle density matrix
directly. The momentum distribution is defined by the Fourier
transform with respect to $i-j$ of the one-particle density matrix
with the form
\begin{equation}
n(k)=\tfrac{1}{L}\sum^L_{i,j=1}e^{-ik(i-j)}\rho_{ij},
\end{equation}
where $k$ denotes the momentum. The eigenfunctions or natural
orbitals $(\phi^\eta_i)$ of the one-particle density matrix
\cite{Penrose} can be obtained by solving
\begin{equation}
\sum^L_{j=1}\rho_{ij}\phi^\eta_j=\lambda_\eta\phi^\eta_i,
\end{equation}
which can be understood as being effective single particle states
with occupations $\lambda_\eta$. For noninteracting bosons, all the
particles occupy in the lowest natural orbital and bosons are in the
BEC phase at zero temperature, however only the quasi condensation
exists for the 1D hard-core bosons. For hard-core bosons in the
incommensurate optical lattice, the strong incommensurate potential
can destroy the quasi-BEC.

The nonequilibrium dynamical properties of expanding clouds of
hard-core bosons on 1D incommensurate lattice after turning off the
harmonic trap suddenly can be also calculated through the equal time
Green function. With the JWT, we can express the equal time Green
function for the expanding hard-core bosons as
\begin{eqnarray}
G_{ij}(t)&=&\langle\Psi^G_{HCB}(t)|b_ib^\dagger_j|\Psi^G_{HCB}(t)\rangle\\
&=& \langle\Psi^G_F(t)|\prod_{\beta=1}^{i-1}e^{i\pi f^\dagger_\beta
f_\beta}f_if^\dagger_j\prod_{\gamma=1}^{j-1}e^{-i\pi
f^\dagger_\gamma f_\gamma}|\Psi^G_F(t)\rangle,\notag
\end{eqnarray}
where $|\Psi^G_{HCB}(t)\rangle$ is the wave function of the
hard-core bosons at $t$ after turning off the harmonic trap and
$|\Psi^G_F(t)\rangle$ is the corresponding one for the
noninteracting fermions. The wave function $|\Psi^G_F(t)\rangle$ can
be easily calculated with the initial wave function
$|\Psi^G_F\rangle$
\begin{eqnarray}
|\Psi^G_F(t)\rangle=e^{-iH'_Ft}|\Psi^G_F\rangle=\prod_{n=1}^{N}\sum_{l=1}^LP_{ln}(t)f_l^{\dagger
}|0\rangle,
\end{eqnarray}
which is still a product of time-dependent single-particle states,
where $\hbar$ has been set to be unit of time in the evolution
operator, $H'_F$ is the $H_F$ in Eq.(\ref{eqn1}) with $V_H=0$, and
$P(t)$ is the matrix of $|\Psi^G_F(t)\rangle$ in the same way as
$|\Psi^G_F\rangle$. Then we can use the method described above to
calculate the equal time Green function, yet we can get the
one-particle density matrix, density profile, momentum distribution,
natural orbitals and their occupations.

\begin{figure}[tbp]
\includegraphics[width=\linewidth, height=10cm, bb=25 20 303 235]{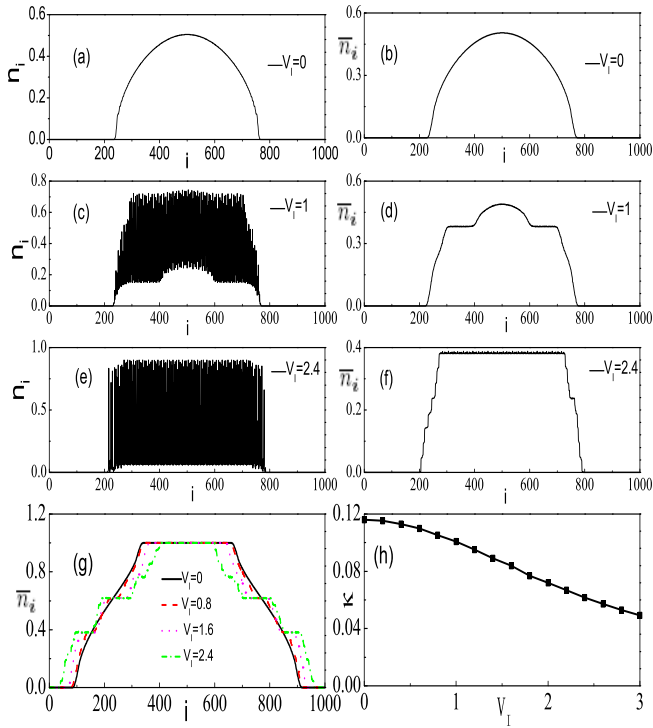}
\caption{(Color online) The density profiles(a,c,e) and the local
average density distributions(b,d,f) for the systems with 1000
sites; 200 bosons; $V_H=3\times10^{-5}$; $\alpha=(\sqrt{5}-1)/2$;
and (a,b)$V_I=0$, (c,d)$V_I=1$, and (e,f)$V_I=2.4$. (g): The local
average density distributions for the systems with 1000 sites, 600
bosons, $V_H=3\times10^{-5}$, $\alpha=(\sqrt{5}-1)/2$ and different
$V_I$. (h): The mean particle fluctuation vs $V_I$ for the systems
with 1000 sites, 200 bosons, $V_H=3\times10^{-5}$,
$\alpha=(\sqrt{5}-1)/2$.} \label{Fig1}
\end{figure}

\section{STATIC PROPERTIES OF HARD-CORE BOSONS CONFINED IN HARMONIC TRAP}

Hard-core bosons in harmonic trap without the incommensurate
potential have been studied by Rigol and Muramatsu \cite{Rigol} in
detail. Quantitatively they characterize the system in a harmonic
trap with the length scale set by combination lattice-confining
potential $\zeta=(V_H/t)^{1/2}$, and the associated characteristic
density $\widetilde{\rho}=N/\zeta$ \cite{Rigol1}. They found that
there is a critical characteristic density
($\widetilde{\rho}_c\sim2.6-2.7$) in the system, for
$\widetilde{\rho}<\widetilde{\rho}_c$ the whole system is in the
superfluid phase at zero temperature, and for
$\widetilde{\rho}>\widetilde{\rho}_c$ there is a phase separation in
the system with a Mott insulating plateau in the middle of the trap
with filling factor equaling one surrounding by the superfluid phase
on the two sides. In this paper we mainly study the influence of the
incommensurate potential which acts as the quasi-random potential
and leads to localization of particles. We focus our study on
systems with low characteristic density
($\widetilde{\rho}<\widetilde{\rho}_c$). For system with high
characteristic density the properties are similar to the one with
low characteristic density except that there is a Mott insulator
plateau in the system. The Mott insulator plateau is not
dramatically influenced by the incommensurate potential because the
particles in the Mott phase are already localized with one-particle
density matrix $\rho_{ij}=\delta_{ij}$. Without especially
illustration, properties studied in this section are for systems
with low characteristic density.

Following the method described in the above section, we can get the
one-particle density matrix. The density profile is given by the
diagonal elements of the one-particle density matrix, i.e.,
$n_i=\rho_{ii}$. In the low characteristic density region, the
density profiles for three different strength of the incommensurate
potential are shown in Fig.\ref{Fig1}. In the absence of the
incommensurate potential ($V_I=0$), the density profile is shown in
Fig.1(a) with all the bosons being in the superfluid phase at the
vale of the harmonic trap. As $V_I$ increases but is still small,
the density profiles basically have the arc shape, but there are a
lot of drastic oscillations in the profiles induced by the
incommensurate potential. The amplitude of the oscillations becomes
more and more large with the increase in $V_I$. When $V_I$ becomes
large enough, the density almost oscillates in the range of $0$ to
$1$. The density profile looks like a belt at the center of harmonic
trap. Additionally, the width of the density profile almost does not
change with the change of $V_I$. For system in high characteristic
density region, the Mott plateau in the density profile basically
does not change with the increase in $V_I$ because the particles in
this phase are already localized. However the superfluid regions
exhibit similar behavior as the system in low characteristic density
region.

Since there are many drastic oscillations in the density
distributions, it is hard to tell that in which area the density is
high in a density profile. In order to reduce the drastic
oscillations, we define the local average density $\overline{n}_i$
as
\begin{eqnarray}
\overline{n}_i=\sum_{j=-M}^Mn_{i+j}/(2M+1),
\end{eqnarray}
where $2M+1$ is the length to count the local average density with
$M\ll L$. In this work, we set $M=10$. The local average density
distributions for three different values of $V_I$ are also shown in
Fig.\ref{Fig1}. When $V_I=0$, the local average density distribution
is almost the same as the density profile with no drastic
oscillations. As $V_I$ increases, plateaus appear at the shoulders
of the arc (see Fig.\ref{Fig1}(d)), then become wider and wider.
Such kind of plateau is induced by the disorder, and thus we call it
Anderson plateau hereafter. When $V_I>2$, the shape of the local
average density does not change with the increase in $V_I$. In
Fig.\ref{Fig1} (e) and (f), the density distribution and the local
average density distribution for $V_I=2.4$ are displayed.
The number of the plateaus and their locations are related to the
particle number and the strength of the harmonic trap. In
Fig.\ref{Fig1} there happened to be only two plateaus. For system
with high characteristic density, the local average density
distributions for different $V_I$ are shown in Fig.\ref{Fig1}(g).
The Mott insulator plateau always exists in the trap center as $V_I$
increases, except that the edges of the plateau are more and more
dissolved into superfluid phase when $V_I<2$. Meanwhile new plateaus
appear at the two sides of the distribution, which is similar to the
one with low characteristic density as $V_I$ increases. The Mott
plateau in the trap center is characterized by $n_i=1$ which does
not oscillate against the particle density in the Anderson plateaus
due to the existence of the incommensurate potential.

The mean particle fluctuation for various value of $V_I$ is shown in
Fig.\ref{Fig1} (h). It is defined by $\kappa=\sum_{i=1}^L\kappa_i/L$
with $\kappa_i$ being the local particle fluctuation
$\kappa_i\equiv\langle\hat{n}^2_i\rangle-n^2_i=n_i-n^2_i$ for
hard-core bosons. We can see that $\kappa$ decreases as $V_I$
increases because of the incommensurate lattice making the bosons
difficult to hop. For system in high characteristic density region,
the local particle fluctuation for particles in the Mott insulator
plateau always equals zero as $n_i=1$ for any $V_I$, whereas the
particle fluctuation is not equal to zero for particles in Anderson
plateaus. So we can distinguish the Mott region from the superfluid
region or Bose-glass region by the local particle fluctuation.

\begin{figure}[tbp]
\includegraphics[width=\linewidth, height=4.5cm, bb=25 20 303 235]{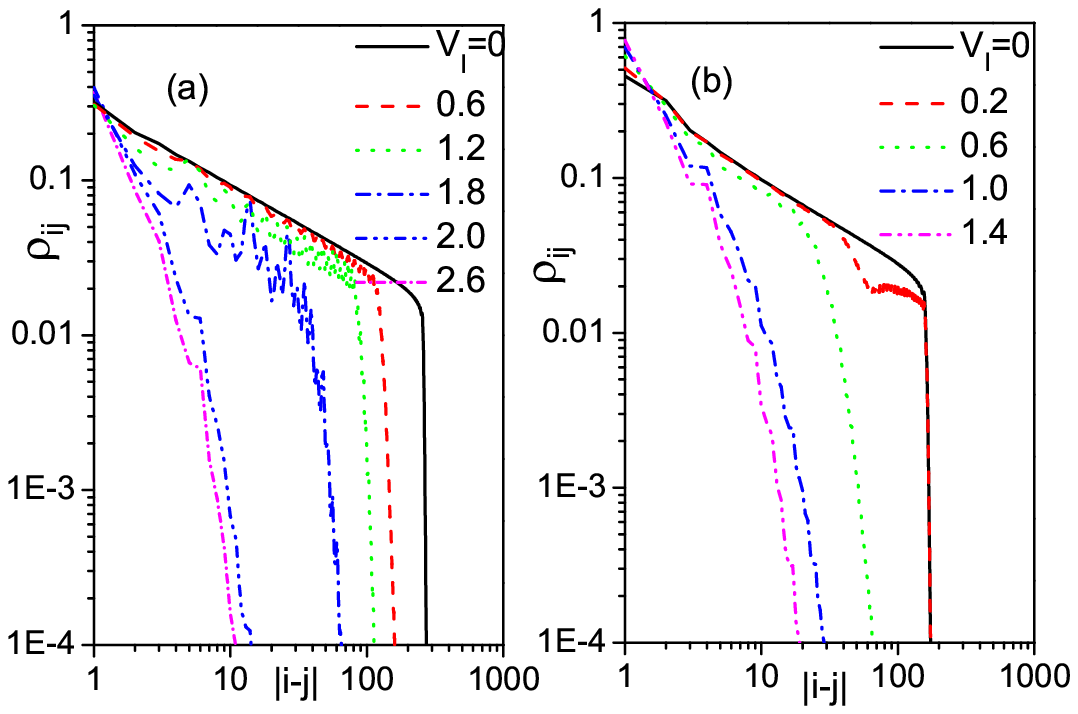}
\includegraphics[width=5cm, height=4.5cm, bb=25 20 303 235]{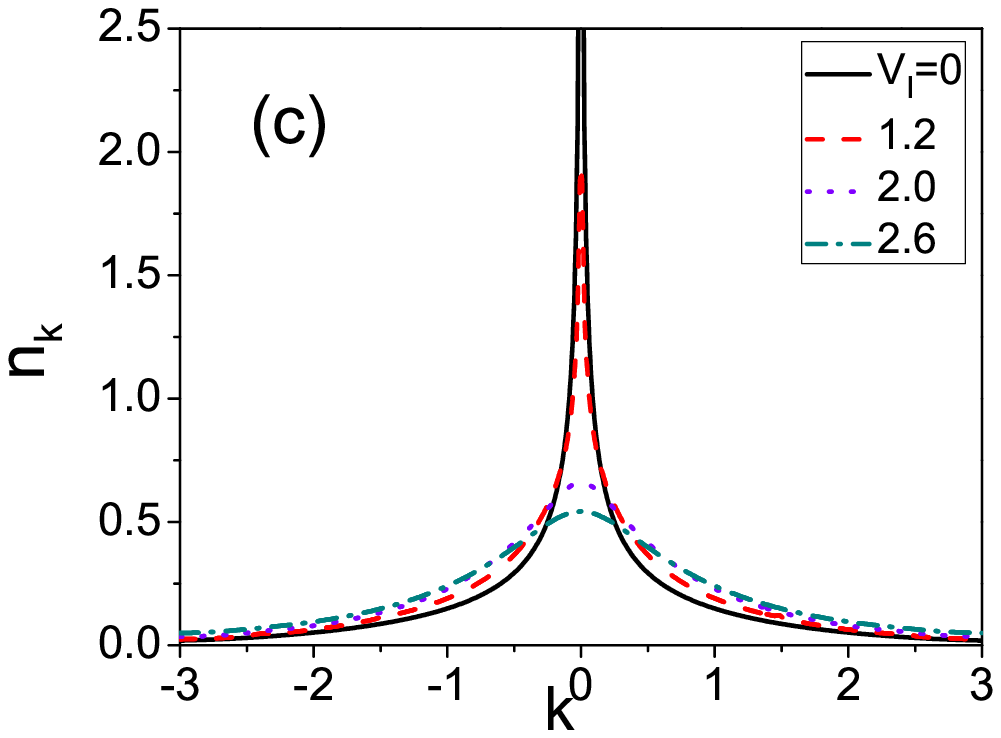}
\caption{(Color online) One-particle density matrices for systems
with 1000 sites; 200 bosons; $V_H=3\times10^{-5}$;
$\alpha=(\sqrt{5}-1)/2$; and (a)$i=501$, (b)$i=601$. (c): The
momentum distributions for systems with 1000 sites, 200 bosons,
$V_H=3\times10^{-5}$, $\alpha=(\sqrt{5}-1)/2$.} \label{Fig2}
\end{figure}
%

In Fig.\ref{Fig2}(a,b), we show the one-particle density matrices
for different value of $V_I$ and site. In the absence of the
incommensurate potential ($V_I=0$), the density matrix has a
power-law decay with exponent $-1/2$ \cite{Rigol}. As $V_I$
increases but is still small, the density matrices still have the
power-law decay, but the exponents are smaller than $-1/2$ and have
lots of oscillations induced by the incommensurate potential.
Power-law decay of the density matrix is the character of the system
in superfluid phase, so when $V_I$ is small, the system in  the
superfluid phase. But things change when the Anderson plateaus
appear in the local average density distribution as $V_I$ increases.
The one-particle density matrices for the sites in Anderson plateaus
have an exponential-law decay, while the ones for the sites out of
the Anderson plateaus still have the power-law decay. Different from
the case without the harmonic confining trap \cite{Cai}, the systems
are not in a uniform phase. When $V_I>2$, the local average density
distribution only consists of several Anderson plateaus, and the
density matrix has the exponential-law decay which is the character
of the system in Bose-glass phase.
For system with $\widetilde{\rho}>\widetilde{\rho}_c$, the
one-particle density matrix is given by $\rho_{ij}=\delta_{ij}$ for
$i,j$ in the Mott insulator plateau, and the one for $i,j$ on the
shoulders in the density profile behaves like the density matrix of
the system in low characteristic region. In Fig.\ref{Fig2}(c), we
show the momentum distributions for different value of $V_I$. When
$V_I$ is small, the momentum distribution is similar to the case of
$V_I=0$ and has a sharp peak at $k=0$ reflecting the coherence of
the system. As $V_I$ increases further, the peak becomes more and
more shallow and the distribution spreads wider. When $V_I>2$, there
is almost no obvious peak. The system is in the Bose-glass phase
with all the effective single-particle states being localized
states, and the coherence of the system becomes small.

\begin{figure}[tbp]
\includegraphics[width=\linewidth, height=9cm, bb=25 20 303 235]{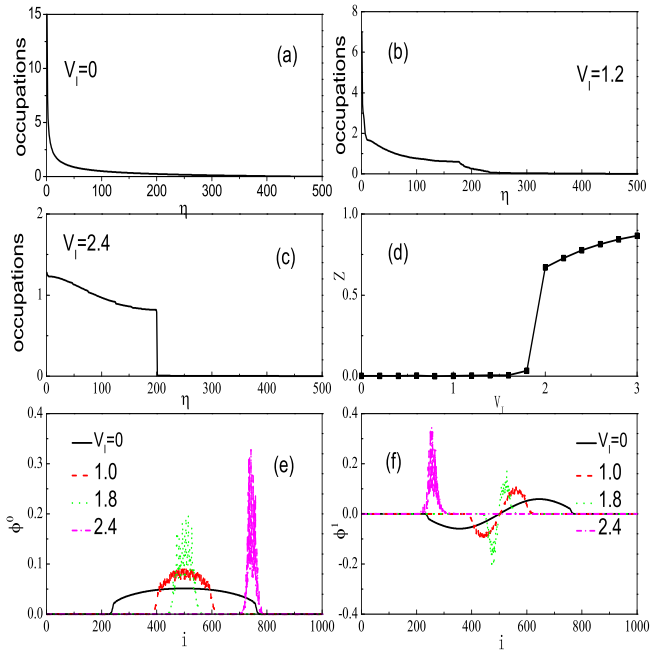}
\caption{Occupations of the natural orbitals for systems with 1000
lattice sites; 200 bosons; $\alpha=(\sqrt{5}-1)/2$;
$V_H=3\times10^{-5}$; and (a)$V=0$, (b)$V=1.2$, and (c)$V=2.4$. (d):
The amplitude of the discontinuation ($Z$) which we define as
$Z=\lambda_N-\lambda_{N+1}$ relates to $V_I$ for systems with 1000
lattice sites, 200 bosons $\alpha=(\sqrt{5}-1)/2$ and
$V_H=3\times10^{-5}$. (e,f): Profiles of the two lowest natural
orbitals for systems with 1000 lattice sites, 200 bosons,
$\alpha=(\sqrt{5}-1)/2$ and $V_H=3\times10^{-5}$.} \label{Fig3}
\end{figure}

Now we study the properties of the natural orbitals and their
occupations for the systems with the low characteristic density. The
occupations of the natural orbitals for systems with different $V_I$
are shown in Fig.\ref{Fig3}. The occupations are plotted versus the
orbital numbers $\eta$, and ordered starting from the highest
occupied one. When $V_I$ is small, the occupation distribution has a
sharp single peak at $\eta=1$ which is the feature of bosons in
contrast to the step function for fermions. With the increase in
$V_I$, the occupation of the lowest natural orbital ($\lambda_1$)
decreases. When the system is in the Bose-glass phase for $V_I>2$,
no an obvious peak appears in the lowest natural orbital.
Additionally we can find that there is a discontinuation in the
occupation distribution at $\eta=N$ when $V_I>2$. We use
$Z=\lambda_N-\lambda_{N+1}$ \cite{Cai} to characterize the
discontinuation for a Bose gas with $N$ particles. In
Fig.\ref{Fig3}(d) $Z$ versus $V_I$ is plotted. There is an obvious
change around $V_I=2$. The difference of the occupations between
systems with the low and high characteristic density is that there
is a plateau in the occupation for system in high $\widetilde{\rho}$
region due to the existence of a Mott insulator plateau. Such a
plateau is basically  not changed as $V_I$ increases. In
Fig.\ref{Fig3} we also show the effect of the incommensurate
potential on the natural orbitals. We plot profiles of two lowest
natural orbitals for different value of $V_I$. Without the
incommensurate lattice, the natural orbitals are similar to the ones
of hard-core bosons in the harmonic trap without lattice. As $V_I$
increases but is still small, the natural orbitals have similar
shapes as the case of $V_I=0$ ones, but the widths of the wave
functions become smaller and smaller due to the particles becoming
more localized.
When $V_I>2$, all the natural orbitals  spread only over a few
lattice sites, and they become typical Anderson localized state
\cite{Cai,Sarma}. The system is in the Bose-glass phase with
exponential-law decay one-particle density matrix
when $V_I>2$. For system with high $\widetilde{\rho}$, the amplitude
of the natural orbitals in the Mott insulator region has to be zero
(see Ref.\cite{Rigol}).

\section{DYNAMICAL PROPERTIES OF HARD-CORE BOSONS}

\begin{figure}[tbp]
\includegraphics[width=\linewidth, height=7cm, bb=25 20 303 235]{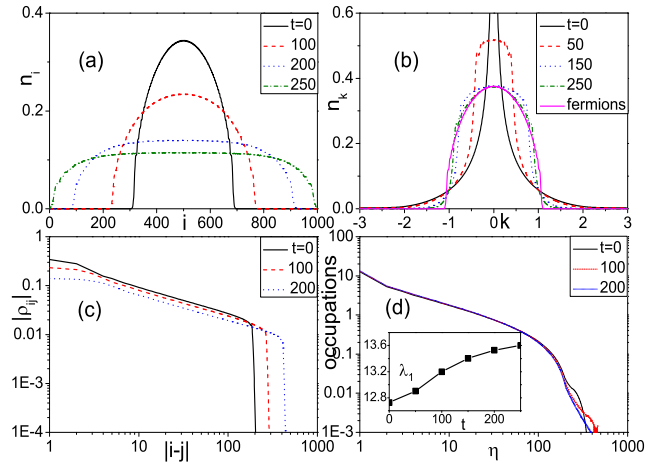}
\caption{(Color online) The evolution of the density profile(a),
momentum distribution(b), one-particle density matrix(c) with
$i=501$, occupations(d) of the natural orbitals for system with 1000
lattice sites, 100 bosons, $\alpha=(\sqrt{5}-1)/2$,
$V_H=3\times10^{-5}$, and $V_I=0$. $t$ is the time after turning off
the trap. The "fermions" curve is the momentum distribution for
system with 1000 lattice sites, 100 free fermions,
$\alpha=(\sqrt{5}-1)/2$, $V_H=3\times10^{-5}$, and $V_I=0$. Insert
of (d): The occupation of the lowest natural orbital vs $t$ for
system with 1000 lattice sites, 100 bosons, $\alpha=(\sqrt{5}-1)/2$,
$V_H=3\times10^{-5}$, and $V_I=0$.} \label{Fig4}
\end{figure}

In this section, we study the nonequilibrium dynamical properties of
expanding clouds of hard-core bosons on 1D incommensurate lattice
after turning off the harmonic trap suddenly. We first consider the
situations that the systems are in the superfluid phase with small
$V_I$ before turning off the trap. The system with $V_I=0$ has been
studied by Rigol and Muramatsu \cite{Rigol1}. In order to see
clearly the effect of a nonzero value of $V_I$ and provide an
example for comparison, we first present results for the case of
$V_I=0$ following \cite{Rigol1}. As shown in Fig.\ref{Fig4}, the
evolution of the density profile is ordinary, i.e., after turning
off the harmonic trap the density profile spreads more and more
wider as the time increases. As for the evolution of the momentum
distribution, shortly after turning off the trap, the peak at $k=0$
disappears. After a long time, the momentum distribution has similar
shape as the distribution of the noninteracting fermions which does
not change during the expansion \cite{Rigol1}. Although the momentum
distribution displays the behavior of fermionization \cite{Rigol1},
the modulus of one-particle density matrix still has the power-law
decay which is the character of the bosons in superfluid phase. Here
we note that the elements of the density matrix are complex numbers
after turning off trap. The occupations of natural orbitals
basically remain unchanged, but the occupations of the lowest
natural orbitals slightly increase during the expansion. This can be
understood as a result due to the increase in the size of the
expanding atomic cloud. The evolution of the natural orbitals has
similar behavior as the density profile which becomes more and more
wider.

\begin{figure}[tbp]
\includegraphics[width=\linewidth, height=10cm, bb=25 20 303 235]{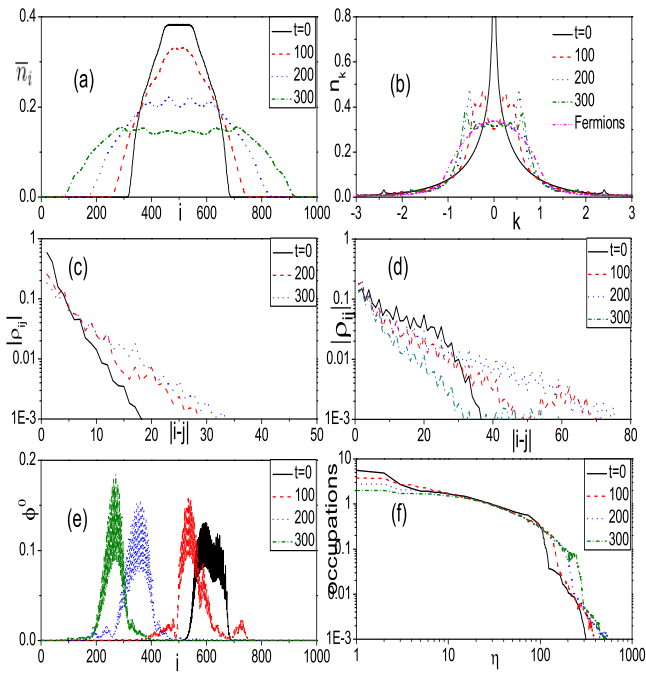}
\caption{(Color online) The evolution of the local average density
distribution(a), momentum distribution(b), one-particle density
matrix with (c)$i=501$, (d)$i=650$, occupations(f) of the natural
orbitals for system with 1000 lattice sites, 100 bosons,
$\alpha=(\sqrt{5}-1)/2$, $V_H=3\times10^{-5}$, and $V_I=1$. (e): The
lowest natural orbitals at different time during expansion for
systems with 1000 lattice sites, 100 bosons,
$\alpha=(\sqrt{5}-1)/2$, $V_H=3\times10^{-5}$, and $V_I=1$.}
\label{Fig5}
\end{figure}

Now we consider the evolution of the system in the presence of the
incommensurate potential. For the case that the system in a weak
incommensurate potential is initially in the superfluid phase, the
evolution of the density profile is similar to the one with $V_I=0$
after turning off the harmonic trap. To give a concrete example, in
Fig.\ref{Fig5} we present the evolution of the local average density
for the system with $V_I=1$. The density profile becomes more and
more wider and the plateaus vanish during expansion. Comparing to
the density profile in Fig.\ref{Fig4}, we can see that the expansion
is much more slow as the incommensurate potential is added in which
acts as a pseudo-random potential and prevents movement of
particles. The evolution of the momentum distribution is shown in
Fig.\ref{Fig5}(b). Shortly after turning off the harmonic trap, the
peak at $k=0$ disappears. In the presence of the incommensurate
potential, there appear some small peaks on the distribution. As
time increases, the peaks of the momentum distribution become small,
and the momentum distribution evolves to the one similar to the
corresponding noninteracting fermions. The evolution of the
one-particle density matrix is shown in Fig.\ref{Fig5}(c,d). The
density matrix $|\rho_{ij}|$ has the exponential-law decay at $t=0$
for $i$ in the Anderson plateaus. For a finite $t$, it still has the
exponential-law decay despite the disappearance of Anderson plateau.
For $i$ out of the Anderson plateaus, $|\rho_{ij}|$ has the
power-law decay at $t=0$ , and it also has the exponential-law decay
after turning off the harmonic trap. In Fig.\ref{Fig5}(f) we show
the lowest natural orbitals for different time during expansion. The
natural orbitals for system at different time have different shapes,
which can be attributed to the scattering of initial wave function
by the incommensurate potential. Correspondingly the occupations of
the natural orbitals are different at different time during
expansion.

\begin{figure}[tbp]
\includegraphics[width=\linewidth, height=4cm, bb=25 20 303 235]{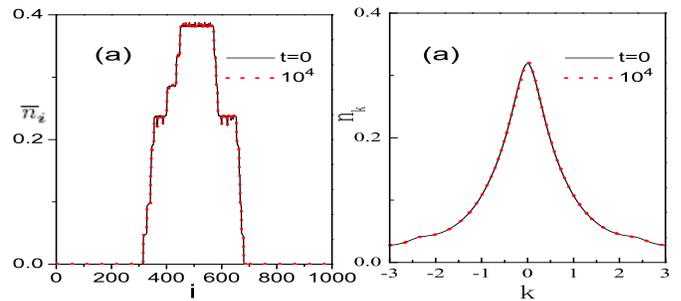}
\caption{(Color online) The evolution of the local average density
distribution(a) and momentum distribution(b) for system with 1000
lattice sites, 100 bosons, $\alpha=(\sqrt{5}-1)/2$,
$V_H=3\times10^{-5}$, and $V_I=2.5$. } \label{Fig6}
\end{figure}

Finally, we study the evolution of system in the BG phase with
$V_I>2$. When $V_I>2$, the system is in the BG phase with all the
effective single-particle states being Anderson-localized states.
After turning off the harmonic trap, the evolution of the system is
very slow, and basically remains unchanged even after a long time. As
an example, we show the evolutions of the local average density and
momentum distribution in Fig.\ref{Fig6}, in which we cannot find the
difference between the initial one and the one after a long time.
This clearly indicates that the system is completely pinned down by
the strong incommensurate potential. The asymmetry of the local average
density is caused by the asymmetry of the incommensurate lattice.

\section{Summary}
In summary, we have studied the properties of hard-core bosons on
incommensurate optical lattices with harmonic confining trap. Using
the Bose-Fermi mapping and the exact numerical method proposed by
Rigol and Muramatsu \cite{Rigol,Rigol1}, we calculate the
one-particle density matrices, momentum distributions, natural
orbitals and their occupations for both the static system and the
dynamic system at different time. Particularly, we exploit the phase
transition from superfluid to the localized BG phase as the strength
of the incommensurate potential increases from weak to strong, and
the nonequilibrium dynamical properties of expanding clouds of
hard-core bosons on the incommensurate lattice after turning off the
harmonic trap suddenly. For the optical lattice with harmonic
confining trap, the density profiles show obvious different
characters for systems with weak, intermediate and strong
incommensurate potential. When the strength of the incommensurate
potential becomes strong enough, Anderson plateaus are founded in
the density distribution. The shape of the local average density
profile shall not change with the increase in the strength of
incommensurate potential if it exceeds a critical value of $V_I=2$.
When the harmonic trap is suddenly switched off, the expansion
dynamics for the systems with $V_I<2$ and with $V_I>2$ also exhibits
quite different behaviors. All of these quantities give clear
signature that there exists a superfluid-to-Bose glass phase
transition in the system when the strength of incommensurate
potential exceeds $V_I=2$. Our study provides an exact static and
dynamic example which unambiguously exhibits the transition from
superfluid-to-Anderson-insulator in the incommensurate optical
lattice.

\begin{acknowledgments}
This work has been supported by NSF of China under Grants
No.10821403 and No.10974234, programs of Chinese Academy of Science,
973 grant No.2010CB922904 and National Program for Basic Research of
MOST.
\end{acknowledgments}

\end{document}